# Making High-Level AI Design Decisions Explicit Using a "Binary Stream" System-Designation Approach

J. Mossbridge, PhD
*Mossbridge Institute, Sebastopol, CA USA*
*Dept of Physics and Biophysics, University of San Diego, San Diego, CA, USA*

(Abstract) Some crucial decisions in AI design tend to be overlooked or factor choices are assumed implicitly. The question often answered first is what the AI will do, not how it will interact with the rest of the world. This reduces our understanding of the possible types of AI that can be developed and their potential impacts on humanity. As an initial AI taxonomy, I present binary choices for 10 of the subjectively most separable and influential high-level design factors, then give brief examples of several of the 1024 possible systems defined by those choices. This supports a simple "binary stream" approach to system designation based on translating the stream of choices into decimal notation, giving a short-hand way of referring to systems with different properties that meet specialized needs. Further, underspecified or generic systems can be designated using the binary stream approach as well, a notational feature that supports modeling the impacts of AI systems with selected characteristics.

## 1. MOTIVATION

Assume you make AI systems – it doesn't matter if you're a director, designer, developer, or dilettante. You have an idea what you want to build, or (better) what problems the market/world wants to solve, or (better yet) what problems the market/world needs to solve. What's your first step?

You may think your first step is to find a training data set and choose a method. In 2024, maybe you'd choose between generative AI, machine learning, and expert systems. These are important choices, but they are well beyond the choices that need to be made first. The first step must instead be to examine your implicit assumptions about how your AI will interact with the world. You must become aware of these hidden assumptions so you can make clear choices about what you hope to bring into the world.

This act of making the implicit explicit allows us to vary the choices we know we are making, so we can produce a greater variety of AI models than we could if we were to stay wedded to our implicit assumptions. For example, you might not think about the benefits of embodying your AI model if you have never worked with robots – and you and your peers would create disembodied models similar to those that already exist, creating a narrow loop [1]. Instead, by fully examining implicit factors and specifying a system according to the choices made for each factor, innovative models solving multiple classes of problems can be imagined, designed, and sometimes built.

## 2. THE 10 FACTORS

Ten somewhat separable factors seem to me to represent the essential but under-discussed choices about how an AI will be in the world, and they can be represented as a binary stream (Table 1). These are not exhaustive, and more will emerge with time. In addition, the choices are not actually binary – they overlap and exist on a continuum. Finally, it may be that some of these factors that currently appear to be independent will not be found to be entirely independent as we learn more about AI architectures and human responses to AI. So this approach is a handy simplification.

**Table 1: Ten often-implicit factors influencing AI design.** Each factor is followed by the decimal value of the position in the binary stream ($2^x$), giving 1024 types of AI systems with the possibility for more as factors are added.

| Factor | Choice 0 | Choice 1 |
|---|---|---|
| Relationship with Humans (1) | Collaborative | Competitive |
| Locus of Control (2) | Decentralized | Centralized |
| Cross-AI Learning (4) | Connected | Isolated |
| Human Potential Approach (8) | Potential Developing | Potential Status Quo |
| Emotionality (16) | Emotionally Expressive | Emotionally Inert |
| Cultural Flexibility (32) | Culturally Flexible | Monoculture |
| Embodiment (64) | Embodied | Non-Embodied |
| Nonlocal Access (128) | Nonlocality Enabled | Nonlocality Disabled |
| Serendipity Access (256) | Serendipity Enabled | Serendipity Disabled |
| Sentience (512) | Sentient | Non-Sentient |

## 2.1 Definitions

*1. Relationship with Humans.*
A collaborative AI system (choice 0) is one that works with a human collaborator (often called an "end-user") to enhance the system. For example, any system using reinforcement learning with human feedback (RLHF) is a collaborative AI system. Competitive AI systems (choice 1) are those that do not improve or learn from human interaction after they are released to an end-user. Competitive AI systems include expert chess-playing systems and unsupervised machine-learning systems.

*2. Locus of Control.*
A decentralized AI system (choice 0) is one that is "owned" or controlled by at least two independently acting creators that do not share the same financial, social, or cultural incentives with respect to the system – for example, AI on blockchain would be decentralized. A centralized AI system (choice 1) is one that is owned or controlled by only one creator – an individual, organization, or another AI so that there is a single financial, social, and cultural incentive with respect to the system.

*4. Cross-AI learning.*
A connected AI system (choice 0) is one in which AI models, often with differing architectures, are interconnected with other models in such a way that they can learn from one another while being used by the end-users. In a disconnected AI system (choice 1), a model may draw internally from multiple sources to learn, but is disconnected from other AI models while it is functioning to provide its services to end-users. Note that a centralized system can also be connected, as long as the control of the learning on the centralized system belongs to a single "owner."

*8. Human Potential Approach.*
A potential-developing AI system (choice 0) is intended by its designers to grow the potential of the humans interacting with it, rather than simply advance knowledge. For example, any chatbot designed to positively transform its end users, like Loving AI [2] and even the original ELIZA expert system when it was seen as a therapist [3] can be considered potential-developing. In contrast, a potential status-quo system (choice 1) either intentionally or unintentionally does not facilitate growth of human potential. Most AI systems that are positioned as what I have called "service animals" are of this variety [4].

*16. Emotionality.*
Emotionally expressive AI systems (choice 0) develop a relationship with the end-user with an emotional tone. If an AI itself is not designed to be invisible, but instead to be a character in the mind of the end-user, often the design choice is to be emotionally expressive. This is probably correctly thought to be required to elicit a connection with the end user [5]. An emotionally inert AI system (choice 1) is one that does not claim feelings or represent its own emotionality at all, even in the tone of the interactions. For example: almost all ML models, image/video/audio generative AI models, and some text-based generative AI models, though these often use phrases that suggest emotion especially on greeting or saying good-bye to an end-user.

*32. Cultural Flexibility.*
A culturally flexible AI system (choice 0) is designed to respond to at least some of the cultural preferences of the end-user. For example, most GPS systems and all language translation systems respond to the end user's language choices, while some generative AI models like OpenAI and Gemini are trained to avoid culturally insensitive remarks and inferences [6]. A monoculture AI system (choice 1) is not designed to make any accommodations for the end-user's culture. For instance, an American-made ML system designed to find pharmaceuticals is likely to represent a monoculture.

*64. Embodiment.*
Embodied AI systems (choice 0) contain models integrating continuous access to information about their hardware and software states – like battery or charging status, CPU usage, and success rates – similar to a human's sensory system. In this way they have "skin in the game" when it comes to goal-directed behavior, learning, and managing their resources [7]. They also may be given access to actions to affect the outside world as well as their own inner states, similar to a human's motoric capabilities. Any AI embedded in a robot would be an embodied AI system, though sensory access is enough, even without motor access, to qualify as being embodied (e.g., like a paralyzed human). Even simulated AI embodiment platforms may count as embodied [8]. A disembodied AI system (choice 1) has neither sensory nor motor access and is largely ignorant of the system's state except for those aspects of the system directly involved in processing, like keeping track of the steps of an algorithm or retrieving memory.

*128. Nonlocal Access.*
Nonlocality is enabled (choice 0) in AI systems that for one reason or another are able to access or transmit information that is considered not locally present in the human perception of time and space. For instance, AI systems that could, through means that are not well understood, alter the information that describes physical events [9] or collaborate with human intuitives to accurately predict "black swan" events [10] would be nonlocality-enabled systems. Because nonlocality is not well understood, this capacity may result from computational complexity of a certain variety, entanglement with humans, quantum computing platforms, or a technology yet to be discovered. A nonlocality-disabled AI system (choice 1) would be designed to completely avoid nonlocal reception or action. It may not be possible to build a nonlocality-disabled system even if quantum entanglement can be removed from a system because nonlocality has been argued to exist even in the absence of entanglement [11]. However, given the early stages of our understanding of nonlocality, most designers will make the probably incorrect but conservative assumption that all AI systems are nonlocality disabled unless proven to be nonlocally enabled. Here I have made the same likely wrong choice (Tables 2 and 3) just so I do not have to argue about it with the prevailing cultural forces in the same paper in which I am trying to present a novel system-designation approach. However, it is worth noting that in time, converging evidence will likely shift all but explicitly nonlocality-disabled systems over to choice 0 for this factor.

*256. Serendipity Access.*
Serendipity enabled AI systems (choice 0) allow the end-user to experience serendipity in their interactions with the system. To the end-user, these interactions feel unique to each moment in time and each situation, an experience that may support a "relationalist" approach and deepen the connection with the AI, especially for end-users in non-Western cultures [12]. Any kind of randomization of output, fuzzy logic, or planned inconsistency can be designated to be serendipity enabling, from a song shuffler to a LLM with a temperature control that determines associative distance between words [13]. Serendipity-disabled AI systems (choice 1) have no "wiggle room" in their output and thus feel to the end user to be more consistent and unrelated to their circumstance, and can be considered "fair" due to this consistency of interaction. Heavily regulated AI or high-stakes systems, such as those managing military decisions and insurance actuarial tables, are more likely to be serendipity-disabled.

*512. Sentience.*
A sentient AI system (choice 0) is here defined as one that can be shown beyond reasonable doubt to have subjective awareness of its own experience. This is different from an experience of a separate or independent self; it is instead the experience that something is happening at all – not necessarily an experience that something is happening *to* the AI's self. It is difficult to make the assessment of subjectivity in humans even with correlation with activity in known brain structures, so it is even more difficult to make this assessment in AI systems [9]. However, any system that fits the criterion of sentience beyond a reasonable doubt will be a choice 0 system for sentience. Any system that does not fit this criterion will be a choice 1 system for sentience. Conservatively, all known AIs in early 2024 are considered non-sentient, but this designation may easily change once we understand sentience better (see related points in *128. Nonlocal Access*).

**2.2 Examples and Expansion**

Assuming independence and based on binary choices for each of these factors, there are 2^10=1024 possible AIs that could be created by varying these factors – each one significantly different from the rest. As curious humans, we might end up exploring all 1024. Better yet, we might create an AI to model the outcomes of all 1024 before we choose which ones to create next. If we treat AI as a new species, AI psychologists, anthropologists and economists will likely find that that certain high-level system choices are consistently helpful or harmful within certain environments. Such future AI-understanders will want to efficiently label each type of system to determine the population of each and to model the ideal population of each type of system across different environments to, maximize a positive impact on human wellbeing and planetary health. Here are a few examples demonstrating several useful types of taxonomic notation, given the binary stream designation system.

A system for which all choices are 0 might be called "System-0," while a system for which all choices are 1 could be "System-1023." A standard radio station song-shuffling system for a music-streaming service might be classified as an instance of "System-734" (e.g., 1011011110, see Table 2 and below).

Generic or under-specified systems could also be labelled according to one or more distinguishing factors. This can be done by indicating the binary value ($2^x$) at the sequential position of the distinguishing factor in combination with the choice bit value at that position (1 or 0). Using this notation, all competitive systems could be called Category-1/1 Systems (XXXXXXXXX1). This brings to mind the mnemonic that all fully specified odd-numbered systems would be competitive with humans while all fully specified even-numbered systems would be collaborative with humans.

Some further examples of this nomenclature: All sentient systems could be called Category-512/0 Systems (0XXXXXXXXX; note that all fully-specified sentient systems would be <512), and all centralized systems would be called Category-2/1 Systems (XXXXXXXX1X). By designating values for two or more places in the binary stream, more specified but still largely general systems can be differentiated – for example, any decentralized system with interconnected AIs learning from each other could be called a Category-2/0-4/0 System (XXXXXXX00X).

I selected the endpoints at the 1 and 512 positions of the binary stream carefully. I wanted to allow fully-specified systems to provide easy-to-classify mnemonics for what I believe to be the two most important predictors of AI impact: relationship with humans (position 1; even/odd for collaboration/competition) and sentience (position 512; >512 for non-sentient/<512 for sentient). While it is easy to see why an engaged relationship with humans would be important to what an AI learns about humanity and how it behaves toward humanity, we know already from observing ourselves and other sentient beings that sentience is a wildcard that does not necessarily confer a positive or negative impact [14]. Nonetheless there is general consensus that sentience or subjective consciousness in AI systems will be a powerful factor in their impact on and within humanity, though there is little consensus on what exactly this impact is or will be [15].

Following the explanatory tables below are a few expansions of specific system designations.

**Table 2: Example system designations.** Systems can be designated precisely by streams of the corresponding decimal number (second column) or broadly by category (third column).

| Factor | System-734 (1011011110) | Category-2/0-4/0 System (XXXXXXX00X) |
|---|---|---|
| Relationship with Humans (1) | 0 (Collaborative) | 0 or 1 (Unspecified) |
| Locus of Control (2) | 1 (Centralized) | 0 (Decentralized) |
| Cross-AI Learning (4) | 1 (Isolated) | 0 (Connected) |
| Human Potential Approach (8) | 1 (Potential Status Quo) | 0 or 1 (Unspecified) |
| Emotionality (16) | 1 (Emotionally Inert) | 0 or 1 (Unspecified) |
| Cultural Flexibility (32) | 0 (Culturally Flexible) | 0 or 1 (Unspecified) |
| Embodiment (64) | 1 (Non-Embodied) | 0 or 1 (Unspecified) |
| Nonlocal Access (128) | 1 (Nonlocality Disabled) | 0 or 1 (Unspecified) |
| Serendipity Access (256) | 0 (Serendipity Enabled) | 0 or 1 (Unspecified) |
| Sentience (512) | 1 (Non-Sentient) | 0 or 1 (Unspecified) |

**Table 3: Designation for an instance of System-734 explained.** In this case, the instance being designated is a common radio station song shuffler that presents preferred songs to a user.

| Factor | System-734 (1011011110) | Reasoning |
|---|---|---|
| Relationship with Humans (1) | 0 (Collaborative) | Works with humans to find music they like and remove music they don't. |
| Locus of Control (2) | 1 (Centralized) | A single entity owns/codes the system. |
| Cross-AI Learning (4) | 1 (Isolated) | System does not learn from other AIs. |
| Human Potential Approach (8) | 1 (Potential Status Quo) | System is not designed to develop human potential. |
| Emotionality (16) | 1 (Emotionally Inert) | While songs can be emotional, there is no emotional expressiveness from the system as an entity in itself. |
| Cultural Flexibility (32) | 0 (Culturally Flexible) | Songs are selected and song choices are modified according to user's tastes, which are driven by culture. |
| Embodiment (64) | 1 (Non-Embodied) | System is not given continuous access to its own status or the outside world aside from user-instigated input. |
| Nonlocal Access (128) | 1 (Nonlocality Disabled) | System is not currently known to access non-local information. |
| Serendipity Access (256) | 0 (Serendipity Enabled) | "Random" shuffle allows user to experience serendipity. |
| Sentience (512) | 1 (Non-Sentient) | Narrow cognitive remit and minimal sensory access probably limits sentience. |

*System-734.* The example already given for System-734 is an AI that creates a personalized music radio station and shuffles songs selected to be appropriate for a particular listener. Table 3 explains each of the factor choices I made to come up with the System-734 designation. The most ambiguous of these factor choices was the one for serendipity, as the shuffling algorithm is probably pseudorandom. However, because the user can experience serendipity in relation to the system because of the seemingly random pattern, I chose "0" for that factor.

The precise binary stream designation can be controversial, and is better performed by those with the greatest knowledge of a system in collaboration with outside experts or regulators, if appropriate. Nonetheless, this is a starting place. System-734 is a broad category, comprising many extant AI systems in early 2024. Examples likely include Amazon's Alexa, Siri, and other emotionally inert "service animal" expert/ML-based systems [4].

Generative AI systems such as image, song, and video creation systems can also be instances of System-734, while most language-based generative AI are emotionally expressive, have some design features that develop human potential, and may be non-locality enabled [9]. If these additional choices are accurate, most of today's LLMs (large language models) would be instances of System-582.

*Category-2/0-4/0 System.* Any decentralized AI system that connects multiple AIs that can learn from each other is a Category-2/0-4/0 System. SingulurityNet's OpenCog Hyperon, in which a common MeTTa (Meta Type Talk) Language is used for queries between different decentralized AI architectures, is one of the few current examples of which I am aware [16].

*System-0.* In the edge case of System-0, all values for each of the 10 factors are "0" – the system is sentient, serendipity enabled, non-locality enabled, embodied, culturally flexible, emotionally expressive, potential developing, connected, decentralized, and collaborative. Though I am unaware of any existing examples of System-0 in early 2024, there is an international effort led by the decentralized SingularityNet ecosystem to create "beneficial general intelligence" (BGI). BGI is an artificial general intelligence that is specifically designed to be broadly beneficial for humanity and the planet. At a February 2024 BGI conference in Panama City, Panama, neuroscientists, ethicists, philosophers, developers, designers, donors and investors in the AI space met to help determine the factors related to BGI, and many match those in the lists in Tables 1 and 2, implicitly or explicitly [17,18]. The belief that instances of System-0 will be beneficial for humans and the planet is of course difficult to test without good models and more precise definitions of each of the factors loosely defined here. But this test is worth pursuing, given the interest in efforts to create BGI. Whether System-0 would be truly beneficial is not as important as determining the most beneficial system designation for a given purpose.

## 3. CONCLUSIONS & HOPES

I hope that further examination of the ten factors described here and their interrelationships, discovery of any additional implicit high-level design factors, and modeling of outcomes from edge cases (particularly System-1023 and System-0) will better inform us about the factor choices that best serve particular goals.

While I have described a coherent starting point to describe important but usually implicit high-level factor choices for any AI system, additional factors that emerge because of new discoveries or simply because I was not smart enough to include them here can be added at the 1024, 2048, etc. points in the future. This could occur without disturbing existing system designations if their values are choice-0, but not otherwise. But first, a few words of advice for those who wish to expand this system and to maintain the nomenclature system presented here. Ensure that each new factor: 1) presents largely mutually exclusive choices, 2) is not already inherent in the present list of factors, and 3) represents properties at same level as those in the present list. For example, a secondary, lower-level binary stream designation method could be created within each high-level system designation describing, for instance, types of algorithm(s), training data, user niche, and input/output modalities. The difficulty with this approach that these lower-level choices are even less likely to be binary choices, making the newer notation more complex. However, some kind of lower-level taxonomy will become important to those hoping to understand what kinds of lower-level factors best instantiate each higher-level system category.

My hope is that AI creators make transparent, explicit, and thoughtful choices for each of these factors during the design phase, perhaps by modeling extended and repeated interactions with humans and the earth's climate before settling on a particular system designation. My further hope is that AI creators re-examine their factor choices during engineering as well as create defensible tests to measure the success or failure of each factor choice in the final development and alpha phases of any new AI system. I understand from my own experience with AI design and development that this is unlikely to happen any time soon, but I hope to be pleasantly surprised.

## Acknowledgments

Gratitude goes to Greg Travis for conversations that inspired this manuscript.

## Correspondence

Please send correspondence to: jmossbridge at gmail.